\documentclass{aa}  

\usepackage{url}
\usepackage{graphicx}
\usepackage{soul}
\usepackage{float} 
\usepackage{placeins}
\usepackage{txfonts}
\usepackage[dvipsnames]{xcolor}
\usepackage{verbatim}
\usepackage{url}
\usepackage{hyperref}
\begin{document} 

   \title{Multi-point study of the energy release and impulsive CME dynamics in an eruptive C7 flare}

  \author{Jonas Saqri\inst{1} \and
      Astrid M. Veronig\inst{1} \and
      Ewan C. M. Dickson\inst{1,2} \and
            Tatiana Podladchikova\inst{3}\and
                 Alexander Warmuth\inst{4} \and 
                Hualin Xiao\inst{2}\and
    Dale E. Gary\inst{5}\and
  Andrea Francesco Battaglia\inst{2,6} \and
    S\"am Krucker\inst{2,7}
     }

   \institute{
                 Institute of Physics, University of Graz, A-8010 Graz, Austria
             \and
             University of Applied Sciences and Arts Northwestern Switzerland, Bahnhofstrasse 6, 5210 Windisch, Switzerland
                          \and
             Skolkovo Institute of Science and Technology, Bolshoy Boulevard 30, bld. 1, Moscow 121205, Russia
                             \and 
                                  Leibniz-Institut f\"ur Astrophysik Potsdam (AIP), An der Sternwarte 16, D-14482 Potsdam, Germany
                                               \and
             Center for Solar–Terrestrial Research, New Jersey Institute of Technology, Newark, NJ 07102, USA
             \and
             ETH Z\"urich, R\"amistrasse 101, 8092 Z\"urich, Switzerland 
                         \and
             Space Sciences Laboratory, University of California, 7 Gauss Way, 94720 Berkeley, USA
}
\authorrunning{J. Saqri et~al.}

   \date{}
 
  \abstract
   {
   The energy release in eruptive flares and the kinematics of the associated coronal mass ejections (CMEs) are interlinked and require favorable observing positions as both on--disk and off--limb signatures are necessary to characterize these events.   
}
   {
   We combine observations from different vantage points to perform a detailed study of a long duration eruptive C7 class flare that occurred on 17 April 2021 and was partially occulted from Earth view. The dynamics and thermal properties of the flare-related plasma flows, the flaring arcade, and  the energy releases and particle acceleration are studied together with the kinematic evolution of the associated CME in order to place this long duration event in context of previous eruptive flare studies.
   }
   {
   We use data from the Spectrometer-Telescope for Imaging X-rays (STIX) onboard the Solar Orbiter to analyze the spectral characteristics, timing, and spatial distribution of the flare X-ray emission. Data from the Extreme Ultraviolet Imager (EUVI) onboard the Solar TErrestrial RElations Observatory-Ahead (STEREO-A) spacecraft are used for context images as well as to track the ejected plasma close to the Sun. With Atmospheric Imaging Assembly extreme ultraviolet (EUV) images from the Solar Dynamics Observatory, the flare is observed off--limb and differential emission measure maps are reconstructed. The coronagraphs onboard STEREO-A are used to track the CME out to around 8\,$\mathrm{R_{\odot}}$.
   }
   {
       The flare showed hard X-ray (HXR) bursts over the duration of an hour in two phases lasting from 16:04\,UT to 17:05\,UT. During the first phase, a strong increase in emission from hot plasma and impulsive acceleration of the CME was observed. The CME acceleration profile shows a three-part evolution of slow rise, acceleration, and propagation in line with the first STIX HXR burst phase, which is triggered by a rising hot (14\,MK) plasmoid. During the CME acceleration phase, we find signatures of ongoing magnetic reconnection behind the erupting structure, in agreement with the standard eruptive flare scenario. The subsequent HXR bursts that occur about 30\,minutes after the primary CME acceleration show a spectral hardening (from $\delta \approx 7$ to $\delta \approx 4$) but do not correspond to further CME acceleration and chromospheric evaporation. Therefore, the CME-flare feedback relationship may only be of significance within the first 25\,minutes of the event under study, as thereafter the flare and the CME eruption evolve independently of each other.
 
   }
   {}

   \maketitle
\section{Introduction}
Solar flares are the result of the impulsive release of free magnetic energy stored in the solar corona, part of which is used for particle acceleration and to heat the coronal plasma (\citealt{priest2002,fletcher2011}). Flares are often, but not necessarily, accompanied by coronal mass ejections (CMEs), with the CME association rate increasing with flare class (e.g.,\citealt{Yashiro2006,ting2020}). These large-scale structures transport plasma and magnetic fields from the sun into the heliosphere and are responsible for some of the most dramatic changes in space weather, including interplanetary shocks and large geomagnetic storms (\citealt{Gosling1993,koskinen2017}). The energy responsible for ejecting CMEs originates from free magnetic energy stored in the solar corona that is impulsively released through magnetic reconnection (\citealt{priest2002,CMELivreviews2012,green2018}), although additional mechanisms might be significant for smaller events \citep{Zhu2020}.
\par
For many CMEs, the evolution of the erupting magnetic structure can be described by three phases. After an initial slow-rise phase, it is rapidly accelerated within the inner corona and then propagates into the heliosphere with roughly constant velocity \citep{zhang2001}. In interplanetary space, other forces become dominant, in particular the magnetohydrodynamic drag, due to the interaction of the CME with the ambient solar wind streams (\citealt{vrsnak2002,cargill2004,vrsnak2004,vrsnak2013}). The three--part CME evolution corresponds to the preflare, rise, and decay phases of the flare associated with the eruption (\citealt{zhang2001,zhang2004,vrsnak2004,temmer2010}). According to the CSHKP model (\citealt{carmichael1964,sturrock1966,hirayama1974,kp1976}), at the beginning of the eruption, a core field (twisted magnetic flux rope) starts to rise and stretch the overlaying arcade. In the arcade below the rising core field, a current sheet forms and magnetic reconnection occurs, and below the reconnection region, flare loops are formed (see reviews by \citealt{priest2002,janvier2015,green2018}).
\par
In the standard flare scenario, part of the energy liberated by the reconnection process accelerates electrons around the reconnection region. The fraction of electrons that propagates downwards along the magnetic field lines deposit their energy into the ambient plasma due to Coulomb collisions with ambient electrons they encounter in the denser chromosphere. As a byproduct, they also produce hard X-rays (HXRs) due to bremsstrahlung emission that occurs when the fast electrons slow down in the field of the ambient ions. Due to the impulsive energy input by the electron beams, the heated chromospheric plasma then expands into the corona along the magnetic field lines, forming bright flare loops in a process referred to as "chromospheric evaporation" \citep{neupert1968}. The instantaneous energy input by electron beams causes a cumulative increase of hot plasma in the solar corona that emits in SXRs. This is often observed as the so-called Neupert effect, in which the derivative of the SXR emission time profile follows the instantaneous HXR emission profile (\citealt{neupert1968,hudson1991,veronig2005})
\par
\begin{figure}
\includegraphics[width=0.4\textwidth]{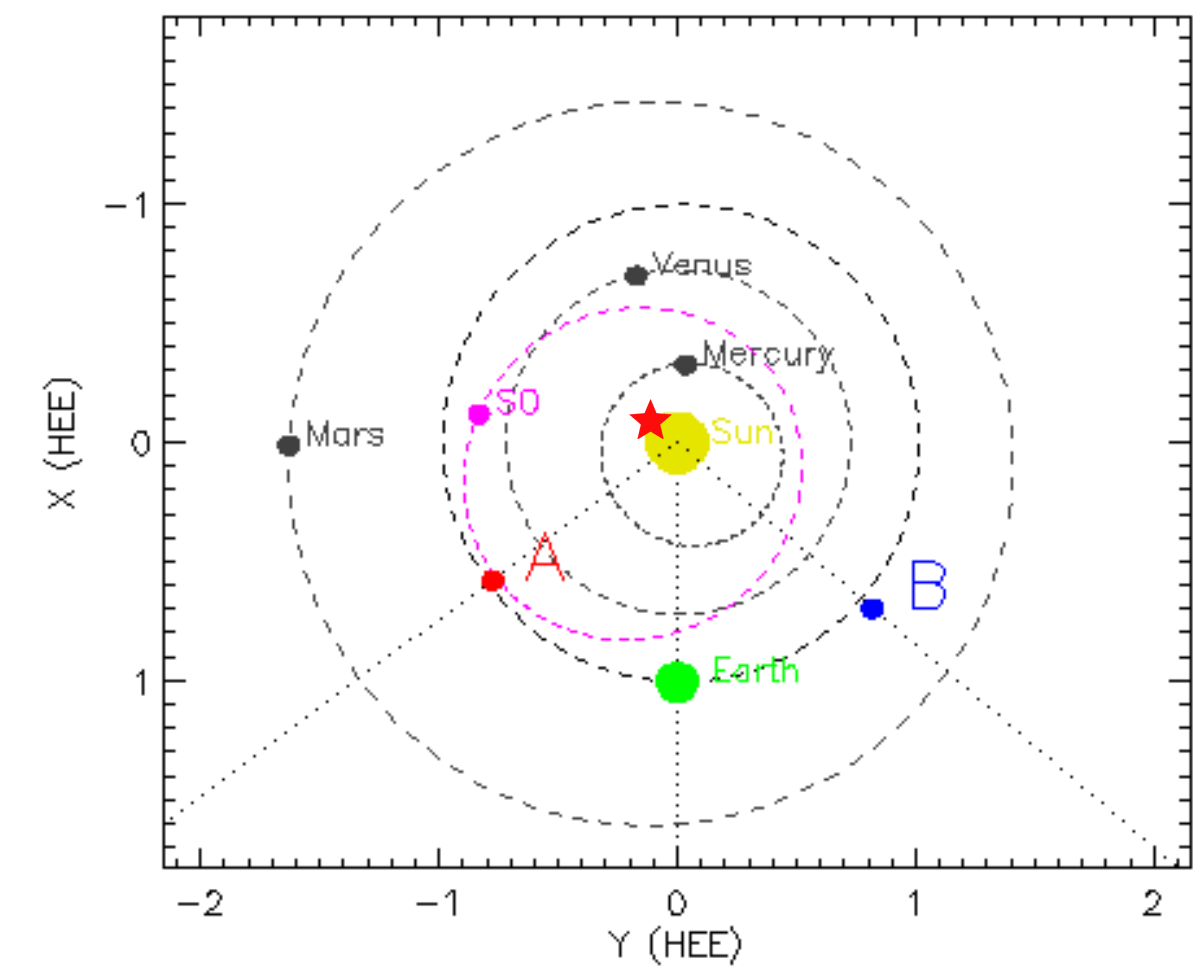}
\caption{Spacecraft positions on 17 April 2021. The flare location is indicated by the red star. Figure adapted from \url{https://stereo-ssc.nascom.nasa.gov/cgi-bin/make_where_gif}.}
\label{f-orbits}
\end{figure}

 While CMEs are best observed in the plane-of-sky (i.e., for source regions close to the limb), many signatures of the associated flare are well observed on the solar disk (e.g., signatures of the flare kernels in UV, EUV, and HXRs). To take advantage of both views, we analyzed a flare with an unusually long duration of nonthermal HXR emission (>1 hr) that occurred slightly behind the limb, as seen from Earth, but that was simultaneously observed on the disk by the Solar Orbiter \citep{mueller2020} and Solar TErrestrial RElations Observatory-Ahead (STEREO-A) spacecraft (see Fig. \ref{f-orbits}). This observational setting allowed us to study in detail the flare energy release and the particle acceleration and response of the atmosphere together with the dynamic evolution of the associated CME and ejected plasmoids.
\section{Data and methods}
The event under study occurred around 16:00\,UT on  17 April 2021 at longitude $\mathrm{\phi}=-110^\circ$ and latitude $\mathrm{\theta}=-18^\circ$ in Stonyhurst heliographic coordinates (i.e., it occurred $20^\circ$ occulted behind the east limb from Earth view).
During the flare, the Solar Orbiter and STEREO-A spacecraft were separated by -$98^\circ$W, $0.4^\circ$N and $-53^\circ$W, $-7.2^\circ$S from the Sun-Earth line, respectively. The Solar Orbiter was at a distance of 0.84\,AU from the Sun, resulting in a 80.2-second light travel time difference relative to Earth. 
\subsection{EUV imaging and white-light coronagraphy}
We studied EUV observations from the Atmospheric Imaging Assembly (AIA; \citealt{lemen2012}) on the Solar Dynamics Observatory (SDO) spacecraft as well as observations from the Extreme Ultraviolet Imager (EUVI; \citealt{howard2008}) on STEREO-A. Data from both instruments were processed to level 1.5 with standard SolarSoftware (SSWIDL) routines. AIA provided images with 0.6\,arcsec pixel resolution and a time cadence of 12 seconds,  while the EUVI images are limited to a cadence of 2.5\,minutes and 1.6\,arcsec pixel resolution. Six coronal AIA EUV filters (94, 131, 171, 193, 211, and 335\,\AA) were used to reconstruct maps of the differential emission measure (DEM) distributions with the regularization inversion code developed by \citet{HK2012}. To perform the DEM inversion, the data were binned by $4\times4$ pixels to improve counting statistics of the faint features under study, resulting in an effective 2.4\,arcsec resolution of the reconstructed maps. For a more detailed description of the DEM reconstruction, we refer to \citet{HK2012} and \citet{saqri2022}.
\par
Observations from the coronagraphs onboard STEREO-A were processed using the standard SSWIDL functionality and running difference images, which were created to enhance the leading edges. With data from the STEREO-A Sun Earth Connection Coronal and Heliospheric Investigation (SECCHI; \citealt{howard2008}) COR1 and COR2 CMEs can be analyzed from about 1.5 to 15\,$\mathrm{R_{\odot}}$. Lower in the corona, the CME structures were followed using the EUVI imagery, which has a field of  view (FOV) of up to 1.7\,$\mathrm{R_{\odot}}$.
\par
To determine the dynamics of the eruption, the leading edge was measured visually three times and the arithmetic mean was used for further analysis. Based on these measurements, we derived the CME kinematics in the corona. To obtain the CME velocity and acceleration profiles, we first smoothed the height-time measurements and derived the first and second direct numerical derivatives. The smoothing technique, described in \cite{podlachikova2017}, when extended to non-equidistant data, optimizes according to two criteria in order to find a balance between data fidelity and smoothness of the approximating curve (see also the applications in \citealt{veronig2018,dissauer2019,gou2020}). Using the derived acceleration profile, we further interpolated to equidistant data points based on the minimization of the second derivatives and reconstructed the corresponding velocity and height profiles by integration. We also obtained the errors of the kinematic profiles by representing the reconstructed CME height, velocity, and acceleration as an explicit function of the original CME height-time data with the standard deviations derived from the three measurements of the leading edge.
\subsection{X-ray and radio observations}
We analyzed data from the Spectrometer-Telescope for Imaging X-rays (STIX; \citealt{krucker2020}). We derived background-subtracted STIX spectra integrated over continuous intervals of 20\,s. Functional fits to the spectrogram data product were done with the OSPEX tool \citep{schwartz2002}, which was recently updated to work with STIX data and is included in the SSWIDL package. For these fits, the sum of an isothermal component (f\_vth) and a thick target model (f\_thick2) were used. For background-subtraction, a five-minute integrated pre-event background spectrum was subtracted from the data. X-ray imaging was performed with the functionality available in SSWIDL \citep{massa2022}. The images were reconstructed with the MEM\textunderscore GE algorithm \citep{massa2020} using a time integration of three\,minutes. To distinguish thermal and nonthermal sources, imaging was performed in the 6-10 and 15-25\,keV energy bins. Since reliable pointing information from the STIX Aspect System \citep{warmuth2020} was not available, due to the comparatively large heliocentric distance, the STIX images were placed manually over the STEREO-A EUVI images. The STIX light curves, spectra, and images were corrected for the light travel time difference relative to Earth of 80.2 seconds.
\par Since a significant part of the flaring plasma was occulted when observed from Earth, GOES flare class was estimated to be C7 based on the relation $\mathrm{log_{10}}(f)=0.622-7.376\,\mathrm{log_{10}}(X^{'})$ of the GOES 1--8\,\AA\, flux $f$ in $\frac{W}{m^{2}}$ and the distance corrected STIX peak 4--10\,keV count rate $X^{'}$ established from 717 previous co-observed events \citep{STIX_data_center}. The XRS instrument onboard the GOES-16 spacecraft was used to calculate the emission measure (EM), and the temperature of the flaring plasma visible to GOES using the filter ratio method for the 0.5--4 and 1--8\,\AA\,channels implemented within the SSWIDL GOES Workbench \citep{goesEMT2005} using one-minute integrated data. In addition, we analyzed background-subtracted cross-power dynamic spectrum data from the Expanded Owens Valley Solar Array (EOVSA) telescope to probe flare signatures in the radio regime from 1--18\,GHz \citep{gary2018}.
\section{Results}
\subsection{Event overview}
While STIX observed the flare against the solar disk, the lower part of the flaring region was occulted by 20$^\circ$ as seen from Earth. Assuming expansion in radial direction, features of the event visible from Earth must therefore be at least 45\,Mm above the solar surface.
\begin{figure}
\centering
\includegraphics[width=0.48\textwidth]{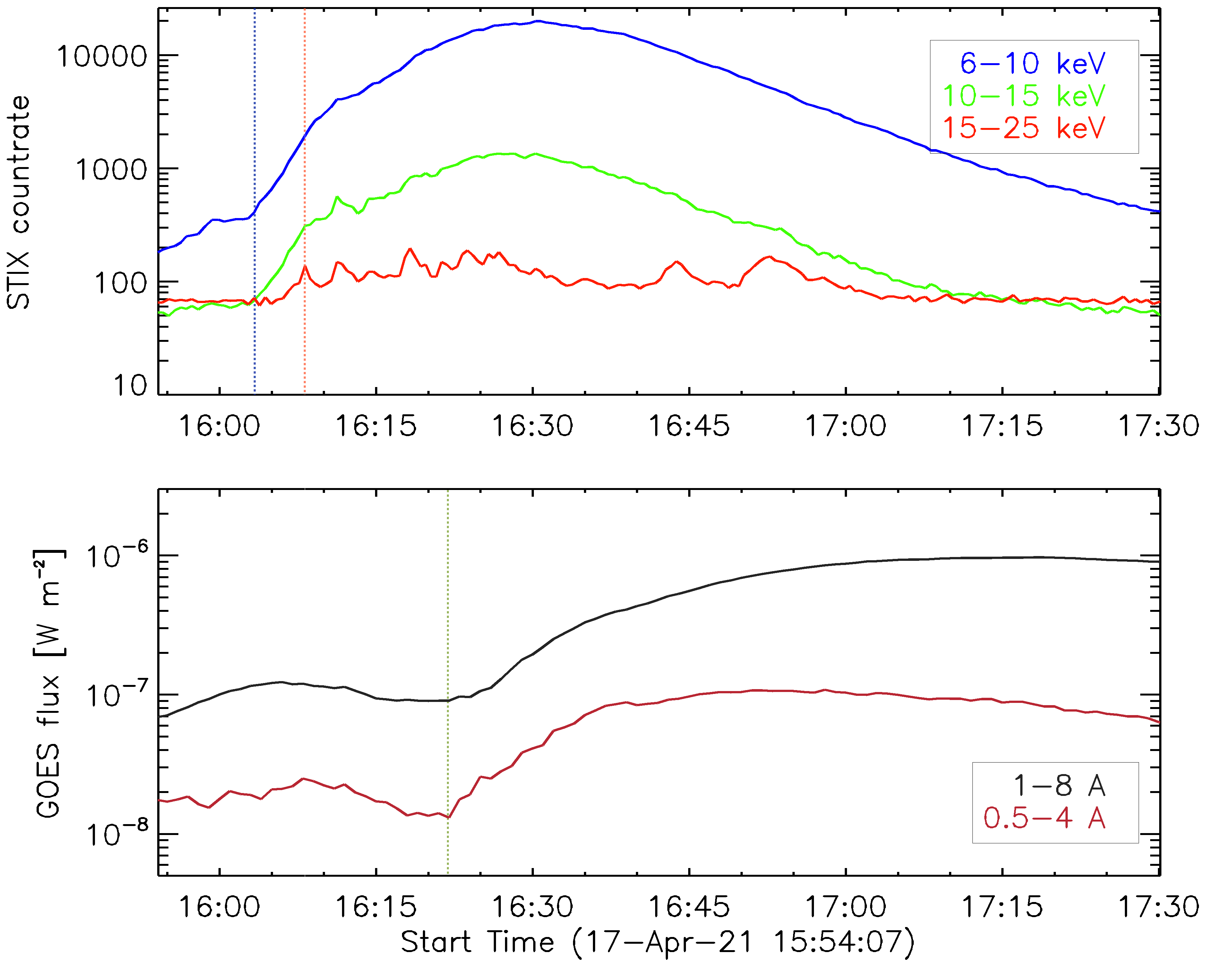}
\caption{Evolution of the flare X-ray emission observed by STIX and GOES. Top panel: STIX observations. STIX HXR light curves in the 6--10 (blue), 10--15 (green) and 15--25\,keV (red) energy bins were derived from spectrogram data smoothed by a 48--sec boxcar averaging. Bottom panel: Light curves of the GOES 0.5--4 (red) and 1--8\,\AA\,(black) SXR channels. The blue vertical line in the top panel marks the start of the impulsive flare phase, and the red line marks the first 15--25\,keV peak. The green line in the bottom panel indicates the time when flare plasma appeared above the limb for GOES.}
\label{f-STIXGOESLQ}
\end{figure}

\begin{figure*}
\centering
\includegraphics[width=0.95\textwidth]{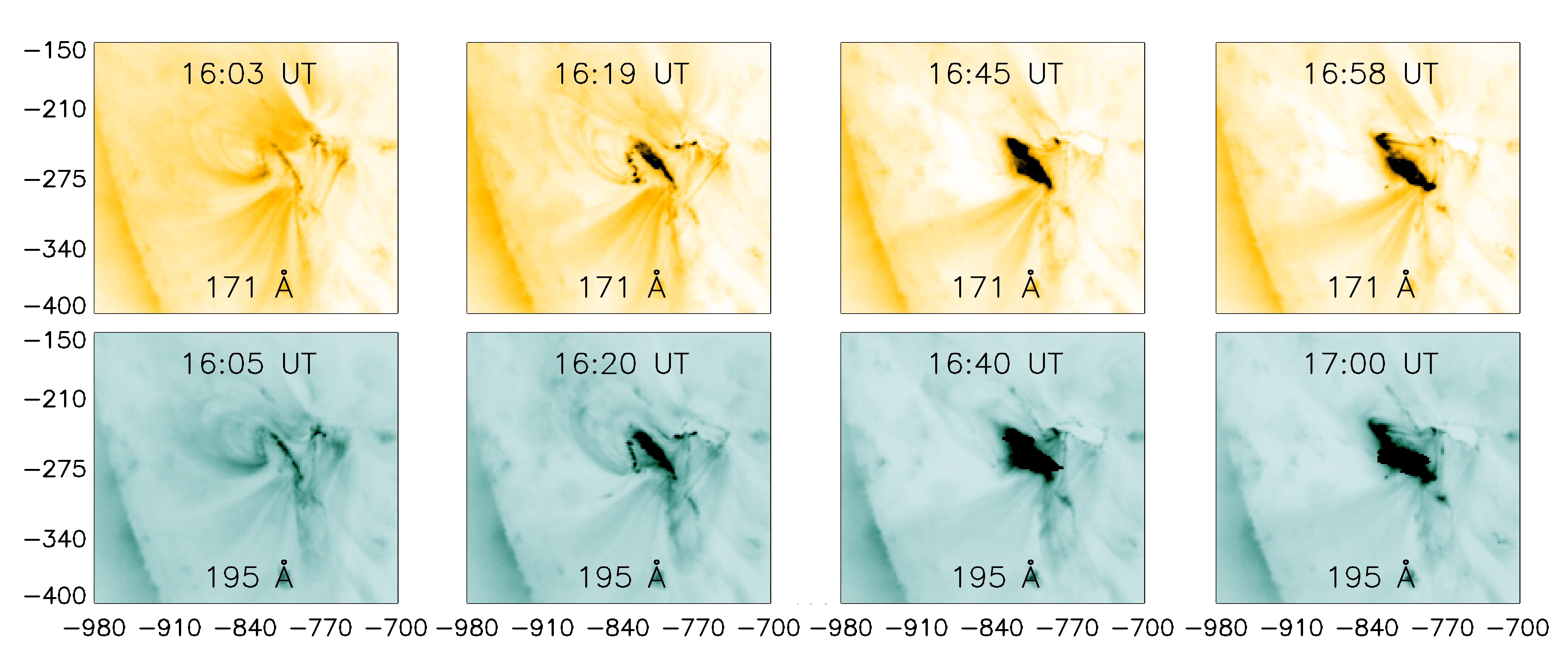}
\caption{Overview of the flare as seen from STEREO-A. Top row: STEREO-A EUVI 171\,\AA\, images. Bottom row: 195\,\AA. Units are given in arcseconds. An animated version is included in the online supplementary electronic material.}
\label{f-EUVI171A}
\end{figure*}

\begin{figure}
\centering
\includegraphics[width=0.48\textwidth]{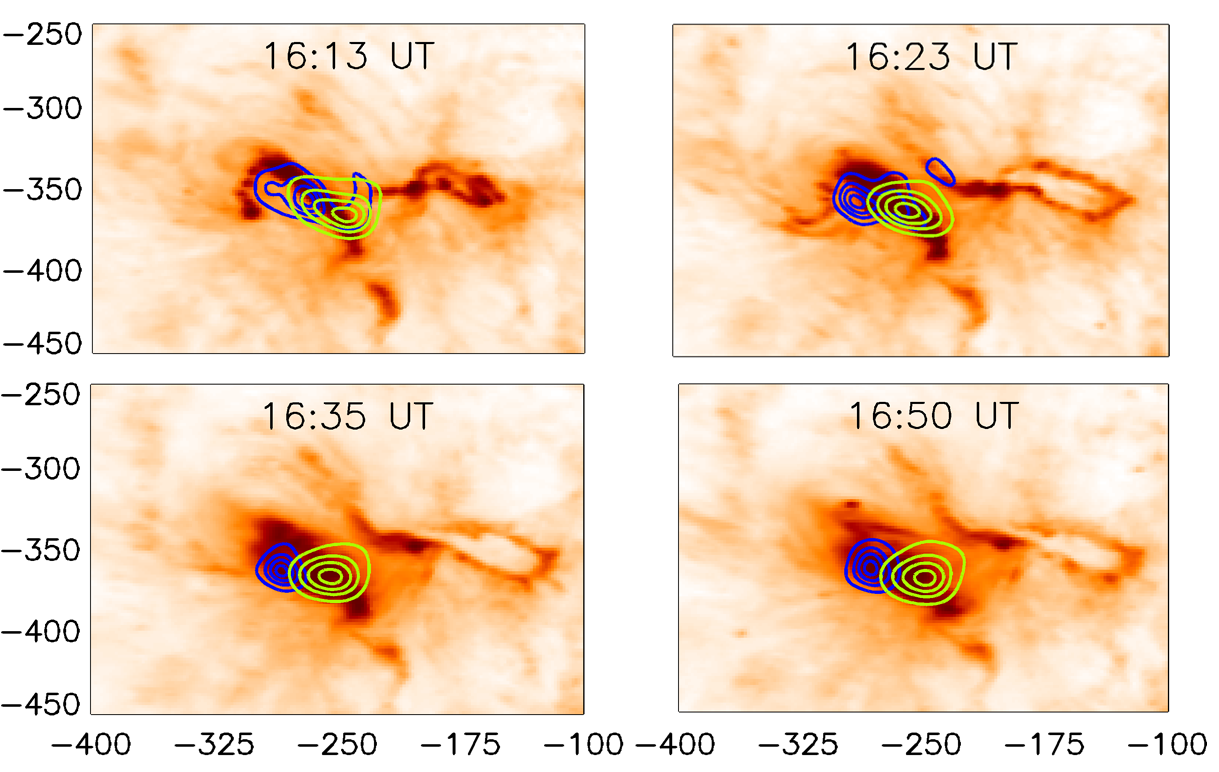}
\caption{Thermal and nonthermal X-ray sources alongside chromospheric EUV observations. Shown in the background are STEREO-A EUVI 304\,\AA\, images transformed to the STIX/Solar Orbiter perspective. Overplotted are contours (30, 50, 70, 90\,\% of the maximum in each image) of STIX spectral images in the 6--10\,keV (green) and 15--25\,keV (blue) energy bands. A movie version can be found in the supplementary online material.}
\label{f-EUVI304ASTIX}
\end{figure}
Figure \ref{f-STIXGOESLQ} shows the STIX HXR light curves in the 6--10, 10--15, and 15--25\,keV energy ranges and the GOES 0.5--4 and 1--8\,\AA\,soft X-ray (SXR) fluxes. The impulsive flare phase starts at 16:04\,UT based on the rise in the STIX 6--10\,keV and 10--15\,keV HXR light curves. The nonthermal energy release of the flare under study lasted from about 16:04\,UT to 17:05\,UT, with several distinct HXR peaks. The first peak in the 15--25\,keV light curve occurred at 16:08\,UT. The first phase of HXR emission lasted until 16:30\,UT and was followed by a phase with two broad peaks starting at 16:40\,UT. The 6--10\,keV bin shows a gradual decay phase with the emission from thermal plasma slowly subsiding after around 16:30\,UT. For both GOES channels, the light curves show an initial enhancement of around 15\,minutes followed by a decrease and then a further increase by an order of magnitude after 16:22\,UT. After 16:22\,UT, some 18 minutes after the start of the event as seen from STIX, more plasma heated by the flare process appeared above the limb (at least 45\,Mm above the solar surface due to the 20$^\circ$ occultation), leading to the substantial increase of the SXR flux. Since the 0.5--4\,\AA\, channel is sensitive to hotter plasma, it peaks earlier and starts to decrease before the 1--8\,\AA\,channel.
\par
The EUV observations from STEREO-A EUVI and SDO AIA also show flaring activity during this time range. Fig. \ref{f-EUVI171A} shows EUVI 171  and 195\,\AA\,maps over the course of the event. The leftmost column shows the onset of the event. The images in the second to left column were taken when the flare was well underway and show the enhanced emission from plasma in the flare loop as well as flare ribbons. Over the next time steps shown, the loop emission in both filters further increases.
\par
Figure \ref{f-EUVI304ASTIX} and the accompanying video show the chromospheric signatures of the flare in the STEREO-A EUVI 304\,\AA\, filter with elongated flare ribbons that expand during the event evolution. Also shown are the results of the STIX spectral imaging in the 6--10  and 15--25\,keV energy bins, which are represented as contours (30, 50, 70, 90\,\% of the maximum in each image). For this comparison, the EUVI 304\,\AA\, images were transformed to the Solar Orbiter observing position. While two extended flare ribbons of enhanced emission can be identified in the EUVI 304\,\AA\, filtergrams, the STIX imaging predominantly recovers nonthermal emission from the eastern flare ribbon. This suggests that the eastern HXR footpoint is substantially brighter or more compact than any other sources regarding the limited dynamic range of the STIX imaging. The 6-10\,keV emission source lies between the two ribbons, indicative of the thermal emission from hot coronal loops connecting the brightest segments of the two EUV ribbons.
\begin{figure*}
\centering
\includegraphics[width=0.95\textwidth]{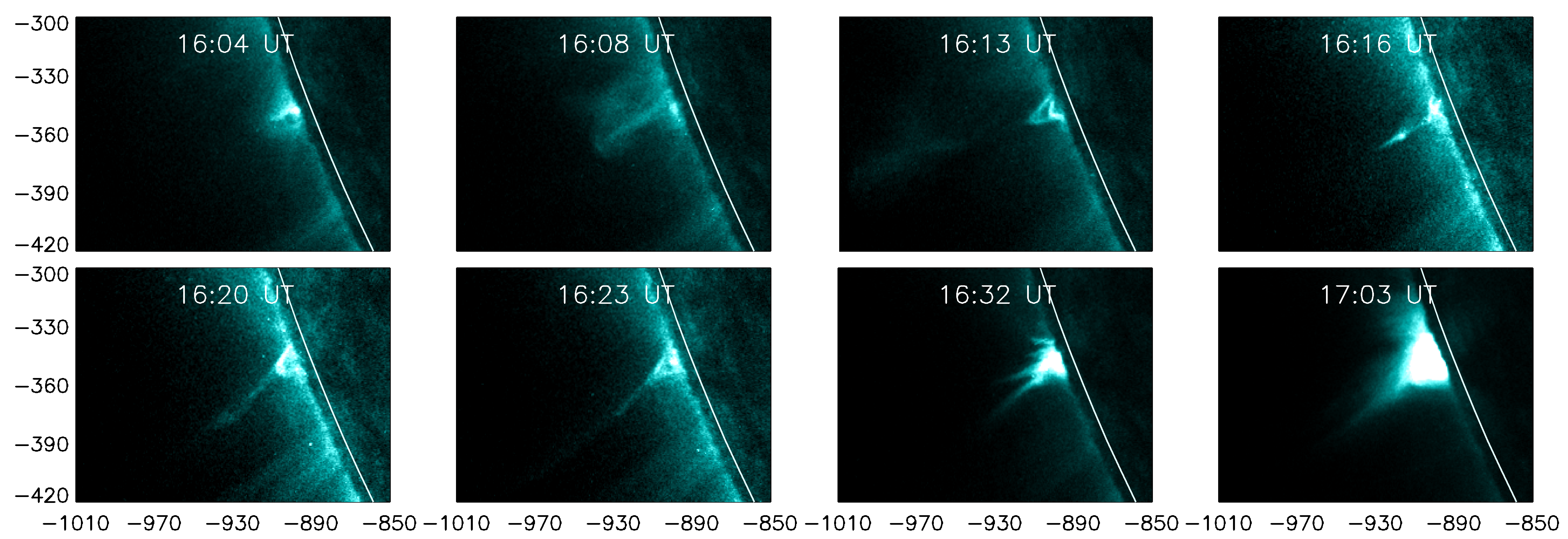}
\caption{Overview of the event seen from Earth in the SDO AIA 131\,\AA\, filter. A movie covering the time span from 15:41 to 17:59\,UT is included in the supplementary online material.}
\label{f-AIA131A}
\end{figure*}
\par
Figure \ref{f-AIA131A} shows the flare evolution in the AIA 131\,\AA\, filter during the one hour of enhanced HXR emission. A video also covering the preflare and a prolonged interval of the decay phase is included in the supplementary online material. As can be seen in the supplementary movie, hot plasma starts to appear above the limb around 15:56\,UT. The frame at 16:04\,UT shows this slowly rising plasmoid. The image taken at 16:08\,UT captures the subsequent eruption of this structure. The appearance of hot plasma above the solar limb in the AIA 131\AA\, filter corresponds to the initial enhancement in the GOES light curves shown in Fig. \ref{f-STIXGOESLQ}. At 16:13\,UT, a cusp structure appears above the limb. Following the initial eruption, smaller instances of upward outflows occur at 16:16\,UT and 16:20\,UT. In the frame taken at 16:23\,UT, another cusp structure with an elongated feature above can be observed, which is indicative of hot plasma associated with the current sheet formed behind the eruption and connected to the hot flare loops. The image taken at 16:32\,UT shows supra-arcade downflows (see e.g., \citealt{shen2022}) that last from 16:28\,UT until 16:50\,UT. Thirty minutes later, shown in the bottom-right panel of Fig.\ref{f-AIA131A}, heated flare plasma fills higher loops and thus becomes visible over the limb. The flare decay phase in AIA 131\,\AA\,is covered for an additional hour in the online movie and shows decreasing emission in the 131\,\AA\, filter sensitive to hot plasma. Due to the occultation of the lower lying parts of the active region when observed from Earth, features visible from Earth must have risen to at least 45\,Mm above the solar surface, which is considerably larger than the distance between the flare ribbons of $\approx$ 25\,Mm estimated from the EUVI observations.
\subsection{Analysis of the erupting plasmoid and CME}
\begin{figure*}
\centering
\includegraphics[width=0.95\textwidth]{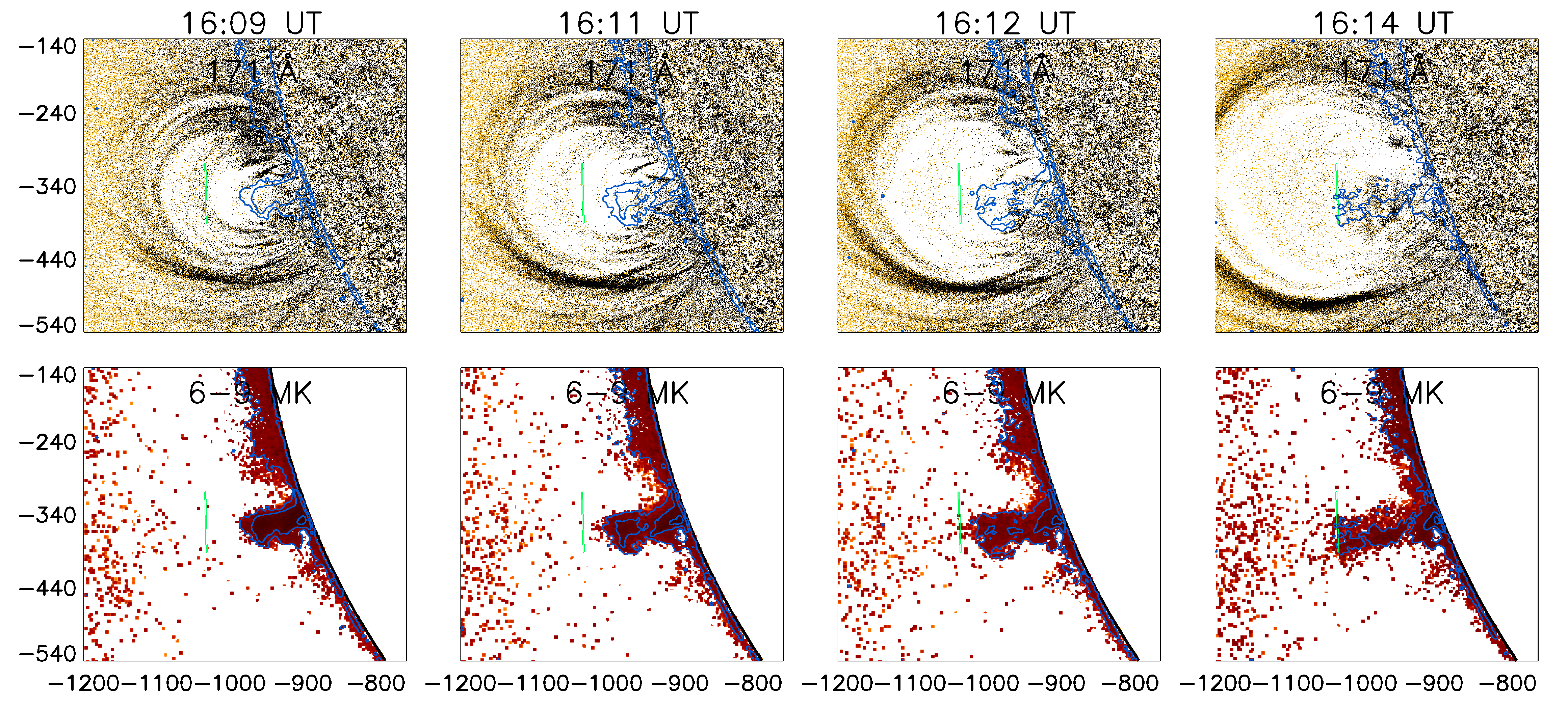}
\caption{Early evolution of the erupting plasmoid and overlying loops. Top row: AIA 171\,\AA\, running difference images created with one-minute difference. Bottom row: EM from 6--9\,MK and median filtered contours in blue. The contours are also overplotted in the top panel. The green vertical line is at a fixed position and is included as a reference to help guide the eye.}
\label{f-plasmoidcomposite171A}
\end{figure*}
As seen from Earth, a rising plasmoid is visible off--limb  during the preflare phase starting at around 15:59\,UT. At 16:08\,UT, at a distance of $\mathrm{0.073 \pm 0.004}$ $\mathrm{R_{\odot}}$ ($\mathrm{50.8 \pm 2.8}$\,$\mathrm{Mm}$) above the solar limb, the plasmoid starts accelerating outwards, and a cusp structure forms beneath it (panels 1-3 of Fig. \ref{f-AIA131A}). This liftoff coincides with the start of the impulsive flare phase, as indicated by the nonthermal STIX HXR emission shown in Fig. \ref{f-STIXGOESLQ}.
\par
The top panels in Figure \ref{f-plasmoidcomposite171A} show the running difference images of the AIA 171\,\AA\,filter created with a one-minute difference. The bottom panels show the EM maps in the 6--9\,MK temperature bin as derived from the DEM analysis using the \citet{HK2012} code. The EM contours (after applying a median filter) are indicated in both rows. As can be seen in the top row, the loops visible in the 171\,\AA\,filter  move up and expand around the outward moving plasmoid. At 16:13\,UT, immediately after the liftoff of the plasmoid, a cusp structure can be observed above the limb (third panel of Fig. \ref{f-AIA131A}).
\begin{figure}
\centering
\includegraphics[width=0.48\textwidth]{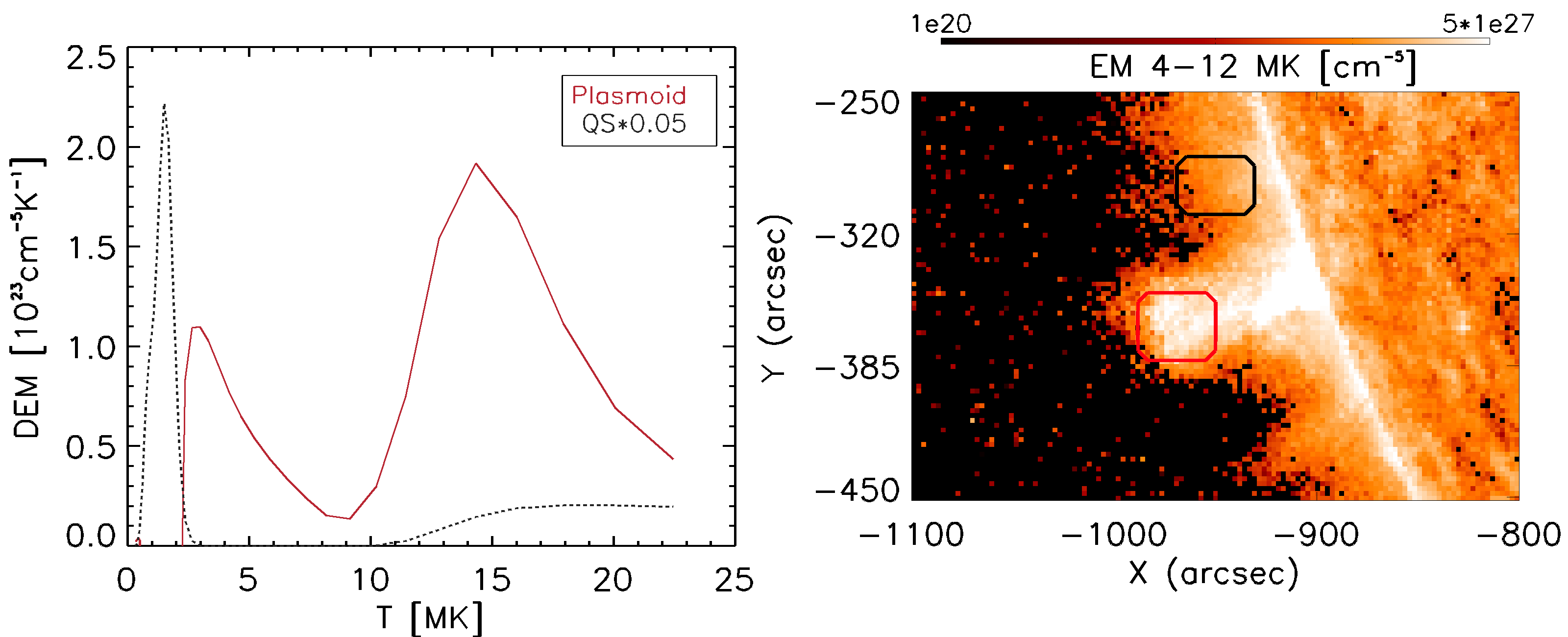}
\caption{DEM analysis of the erupting plasmoid. Right panel: Map of the total EM and regions considered for the DEM. The red box covers the plasmoid, the black box a off--limb quiet Sun region. Left panel: DEM distributions integrated over the regions shown in the right panel at 16:09\,UT.}
\label{f-emMass1}
\end{figure}
\par
\begin{figure}
\centering
\includegraphics[width=0.48\textwidth]{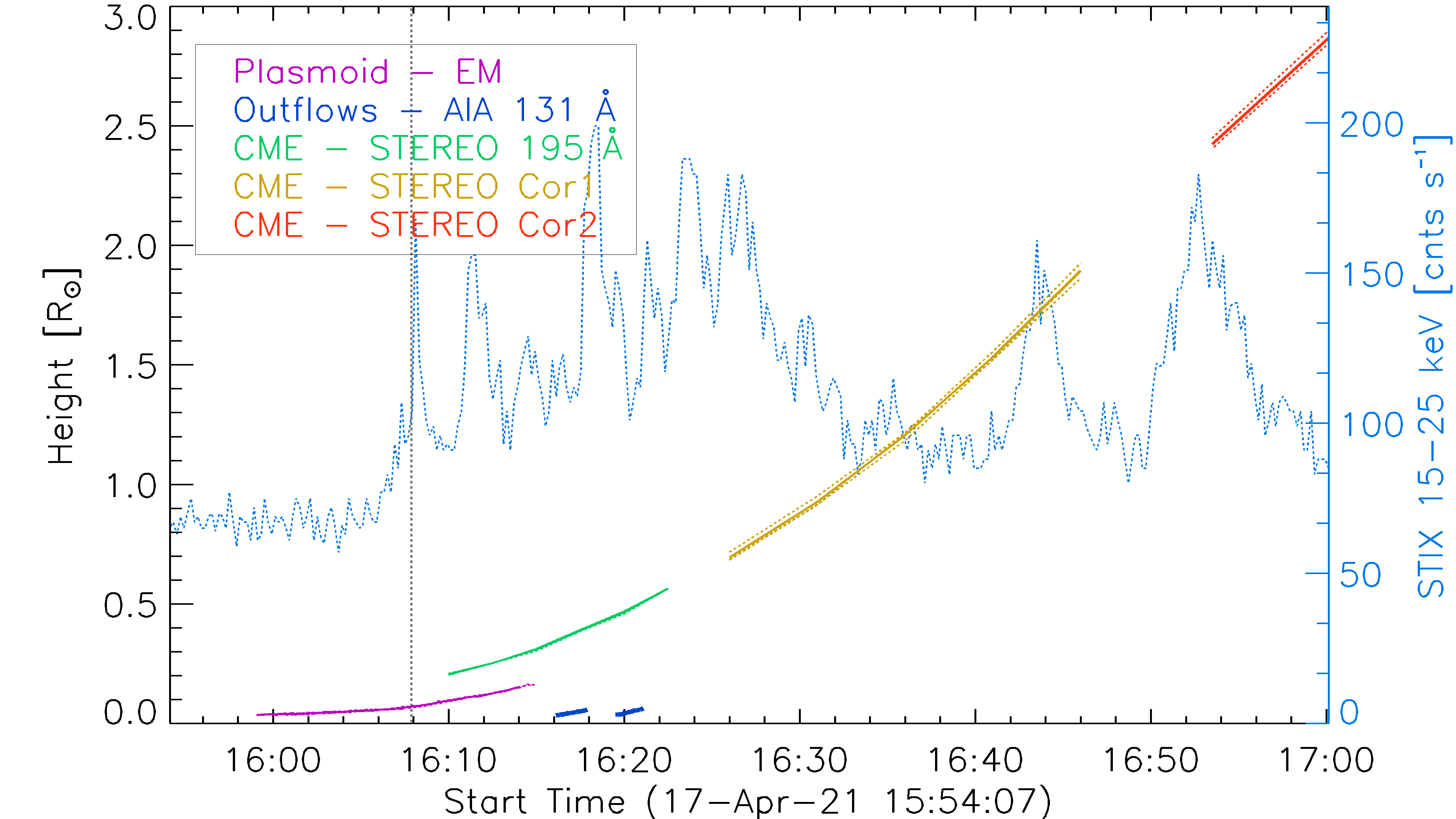}
\caption{Height-time evolution for the erupting plasmoid, the CME front, and the outflows behind the eruption. The blue curve shows the STIX 15--25\,keV HXR light curve. The vertical gray line indicates the time the plasmoid started accelerating.}
\label{f-kinematics0}
\end{figure}

Figure \ref{f-emMass1} shows the pre-event background subtracted DEM distribution totaled over the most intense part of the erupting plasmoid at 16:09\,UT. It is significantly hotter than the ambient coronal plasma, which peaks at 1.5 MK. The DEM profile of the plasmoid shows two peaks, one centered at 3.5\,MK and a larger contribution centered around 14\,MK. We estimated the plasmoid mass to be $\approx 1.5\times10^{11}$\,kg using $\rho=\frac{EM}{h}\times m_{H}$ and $m=\rho \times V$ and by assuming a spherical plasmoid with a diameter equal to the horizontal extent of the red rectangle covering the plasmoid (see right panel of Fig. \ref{f-emMass1}) and the total EM calculated from the DEM shown in the left panel of Fig. \ref{f-emMass1}. The density of the emitting plasma is represented by $\rho,$  and $m_{H}$ is the mass of the hydrogen atom. Since we assumed all the emitting plasma to be located within this sphere, the  line of sight (LOS) depth of the emitting plasma $h$ was set equal to its diameter. The thermal energy contained in the same volume $V$ is estimated to be $$E_{th}=3k_{B}V^{1/2}\displaystyle\sum_{k}^{}T_{k}[\mathrm{DEM}(T_{k}) \Delta T_{k}]^{1/2} \approx 4.7\times 10^{20}\,\mathrm{J},$$ \citep{aschwanden2015}. Using this mass estimate and the velocity estimated from a linear fit to the height-time profile shown in Fig. \ref{f-kinematics0} at 16:09\,UT ($\approx$ 120\,km/s), the kinetic energy of the plasmoid was estimated to be about $1.1\times 10^{21}$\,J (i.e., around 2.3 times larger than its thermal energy). The mass estimate of the ejected plasmoid lies within the lower range of typical CME masses of $10^{11}$ to $10^{13}$\,kg based on white-light observations\citep{vourlidas2002}.
\begin{figure*}
\centering
\includegraphics[width=0.9\textwidth]{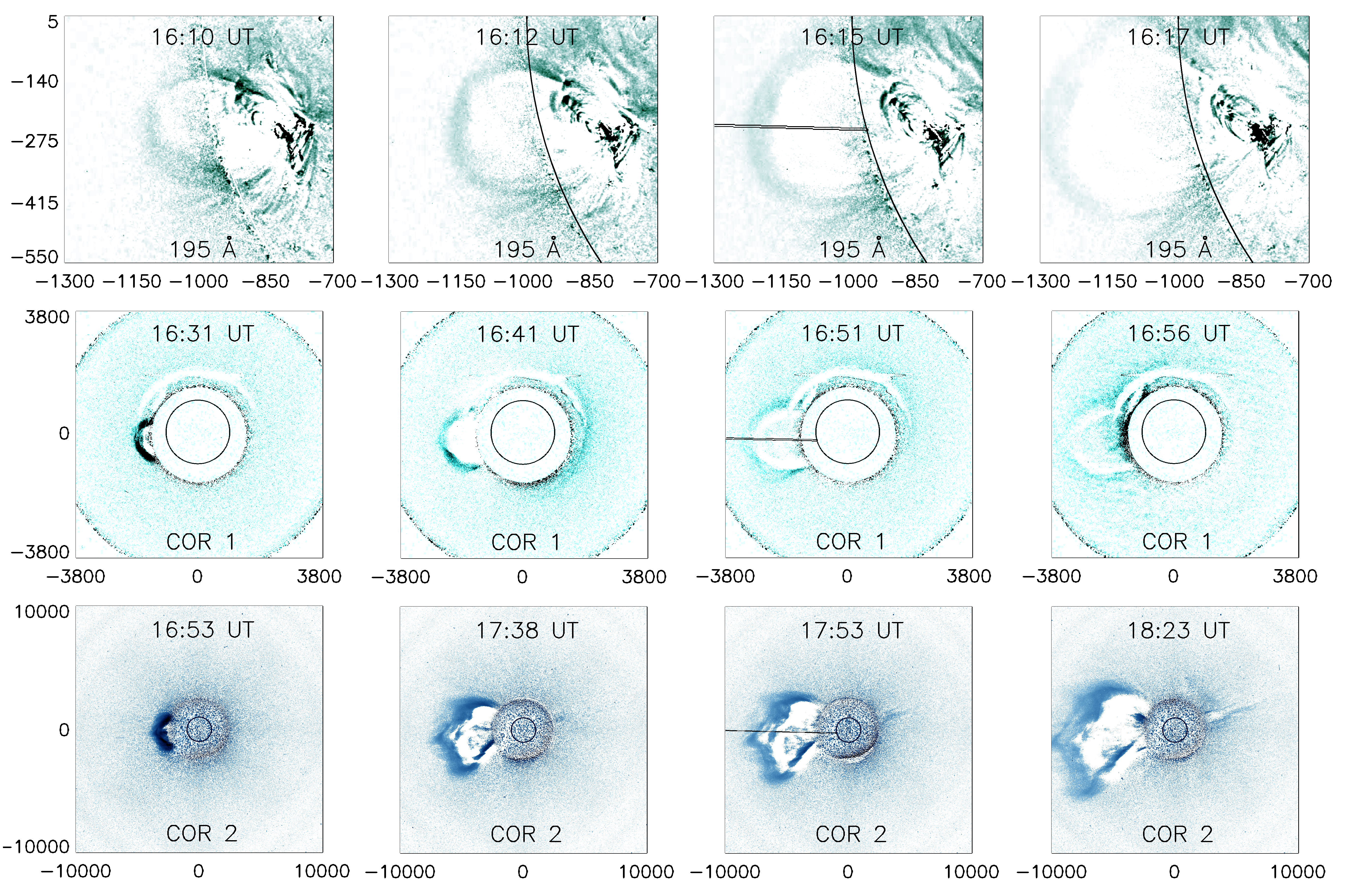}
\caption{Overview of the CME evolution. Top row: STEREO-A EUVI 195\,\AA\, running difference images showing the early stage of the eruption. Middle row: STEREO-A SECCHI COR1. Bottom row: COR2 coronagraph running difference images. Overplotted in black is the direction along which the height measurements were determined. Axis units are in arcseconds.}
\label{f-195WithC23}
\end{figure*}
\par
Figure \ref{f-195WithC23} shows the CME evolution in running difference images from the STEREO-A EUVI 195\,\AA\, filter along with SECCHI/COR1 and COR2 running difference maps. In EUVI 195\,\AA,\,the pileup of plasma accelerating outward is visible as an expanding loop starting at 16:10\,UT, which can then be further tracked from the same viewpoint in the COR1 and COR2 coronagraphs until 18:23\,UT. \par
\par
Figure \ref{f-kinematics0} shows the height-time evolution (measured from the solar limb) for the erupting plasmoid, the CME, and outflows behind the eruption derived from several instruments. For comparison with the impulsive flare evolution, the 15--25\,keV STIX light curve is included in the background.
\par
\begin{figure}
\centering
\includegraphics[width=0.48\textwidth]{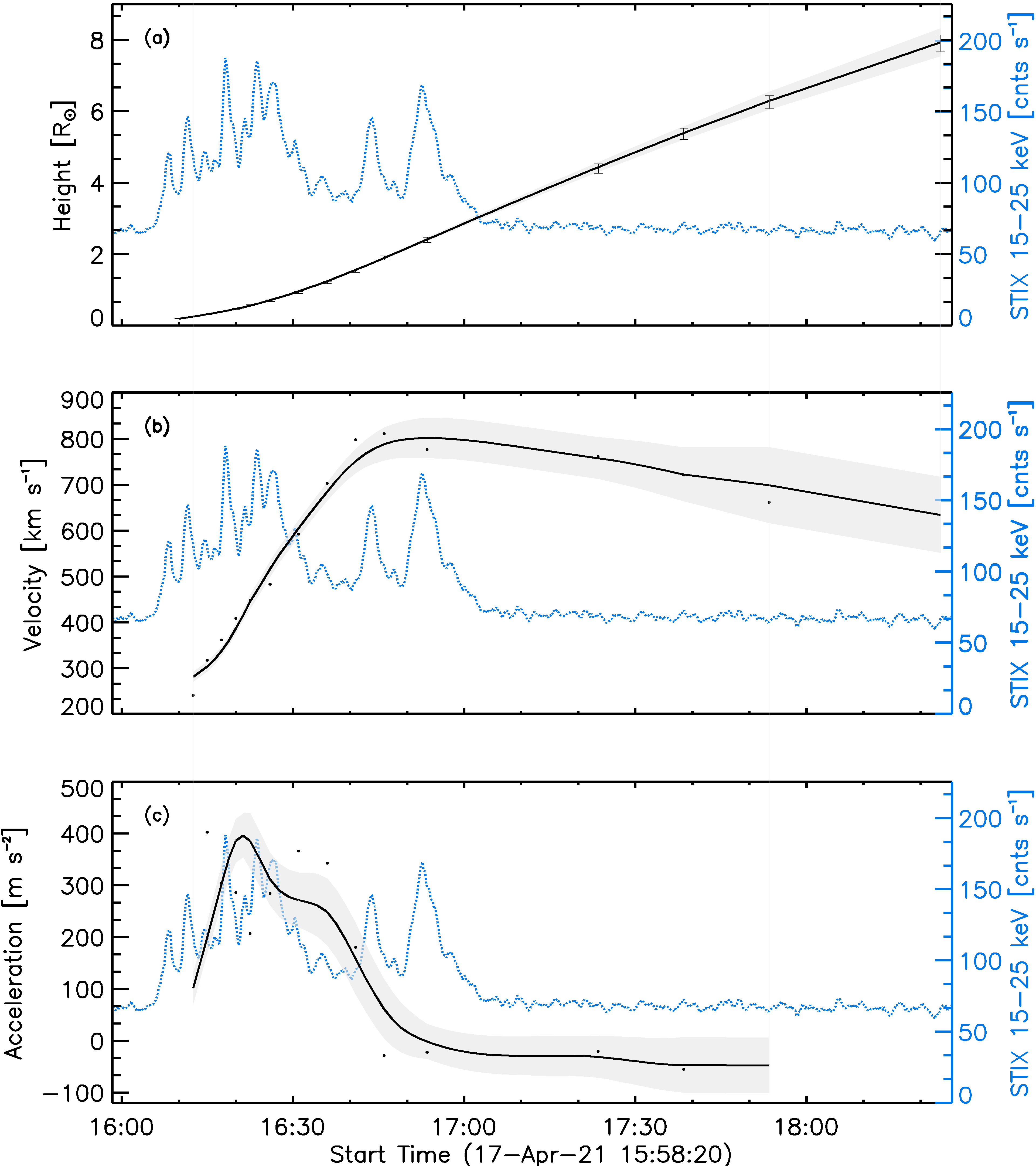}
\caption{CME kinematics derived from combined measurements. Top panel: Distance of the eruption from the solar limb measured with STEREO-A EUVI 195\,\AA\, filter and the COR1 and COR2 coronagraphs. Points with error bars show the measurements and their standard deviation. The black line indicates the smoothed height--time profile. Middle panel: CME velocity obtained from the numerical derivate of the height--time profile. Bottom panel: CME acceleration obtained from the numerical derivate of the velocity profile. The dots in panels b) and c) show the numerical derivatives, the solid lines the smoothed profiles. The shaded areas represent the error ranges. The STIX 15--25\,keV light curve binned over 20\,seconds is shown in blue in the background.}
\label{f-kinematics1}
\end{figure}

Figure \ref{f-kinematics1} shows the evolution of the CME tracked in STEREO-A EUVI 195\,\AA, COR1, and COR2 imagery. From top to bottom, Fig. \ref{f-kinematics1} shows a) the height-time profile (measured from the solar limb), b) the corresponding velocity, and c) acceleration profiles. Also indicated are the CME height-time measurements (a) and first (b) and second (c) numerical derivatives together with error ranges. The profiles were derived using the method of \citet{podlachikova2017}. For comparison with the impulsive flare evolution, the STIX 15--25\,keV HXR light curve is shown in the background.
\par 
After the first HXR peak at 16:08\,UT, which is associated with the erupting plasmoid (Fig. \ref{f-plasmoidcomposite171A}), the impulsive acceleration of the CME lasted from 16:10 until 16:40\,UT, though we note that there was still ongoing acceleration on a smaller level until about 16:50\,UT. The peak acceleration of $\mathrm{395 \pm 42}$\,$\mathrm{m/s^{2}}$ occurred at 16:21\,UT at a height of 0.51\,$\mathrm{R_{\odot}}$. The increase in CME velocity coincides with the first phase of the STIX HXR bursts. After reaching a maximum velocity of $\mathrm{802 \pm 44}$\,$ \mathrm{km/s}$ at 16:54\,UT at a height of 2.4\,$\mathrm{R_{\odot}}$, the CME gradually decelerated to $\mathrm{634 \pm 82}$\,$\mathrm{km/s}$\,until 18:23\,UT. After the end of the main CME acceleration, two HXR peaks (at 16:44 and 16:52\,UT) with intensities comparable to earlier bursts can be observed. During the first of these HXR peaks, the CME velocity was still increasing and reaching its maximum, while the second does not correspond to further CME acceleration (Fig. \ref{f-kinematics1}).
\subsection{EUV reconnection signatures and flare decay phase}
\begin{figure}
\centering
\includegraphics[width=0.48\textwidth]{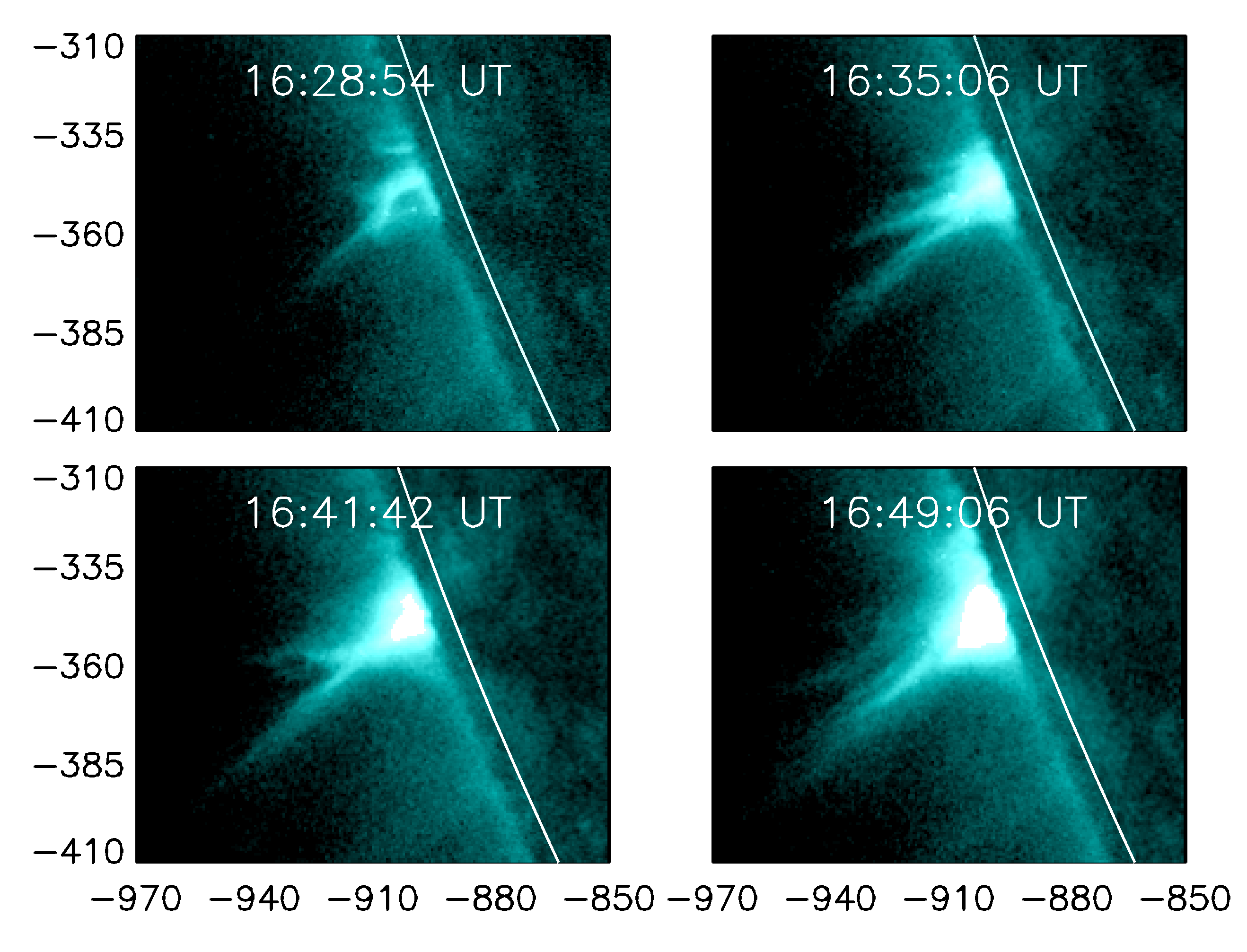}
\caption{AIA 131\,\AA\, filtergrams during the episode of supra-arcade downflows observed from 16:28\,UT to 16:50\,UT. A movie can be found in the supplementary electronic material.}
\label{f-downflows}
\end{figure}

After the CME was launched and the first cusp structure became visible at the limb (16:13\,UT in Fig. \ref{f-AIA131A}), the HXR light curve and AIA EUV observations showed indications of energy release and ongoing magnetic reconnection behind the erupting structure. At 16:16 and 16:20\,UT, two separate upward outflows can be seen in the AIA 131\,\AA\,filter sensitive to hot flare plasma (with peak formation temperature T $\approx$ 20\,MK). The outflows were followed by the appearance of a cusp structure with an elongated feature above the cusp, which is indicative of hot plasma in the current sheet formed beneath the erupting CME. Starting at 16:28\,UT, supra-arcade downflows above the flare loops can be observed until 16:50\,UT. Figure \ref{f-downflows} and the accompanying movie show AIA 131\,\AA\,images over this time span of continuous supra-arcade downflow activity.
\par
\begin{figure}
\centering
\includegraphics[width=0.48\textwidth]{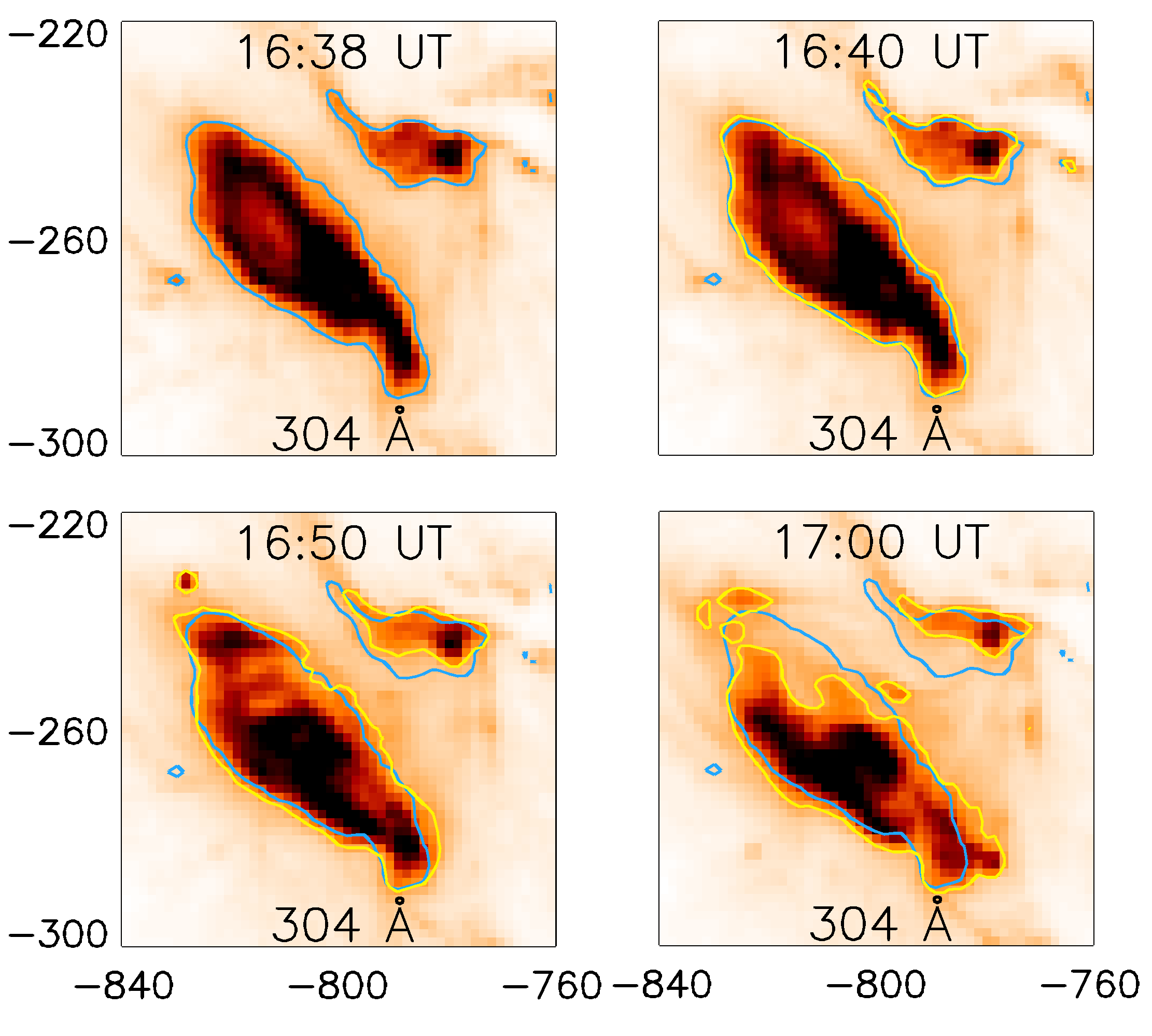}
\caption{EUVI 304\,\AA\, filtergrams showing the flare ribbon evolution after the main CME acceleration phase. Overplotted in yellow are the 20\% instantaneous contours alongside the contours from 16:38\,UT in blue, which are included as reference points to illustrate the movement of the northern flare ribbon.}
\label{f-footpoints}
\end{figure}

Figure \ref{f-footpoints} shows EUVI 304\,\AA\, filtergrams before and during the two last HXR peaks, which occurred around 16:44 and 16:52\,UT, after the main CME acceleration. The 20\% contours of each image are shown alongside the contours from 16:38\,UT. The flare ribbons still move apart slowly, indicating reconnection at increasing atmospheric heights.

\begin{figure}
\centering
\includegraphics[width=0.48\textwidth]{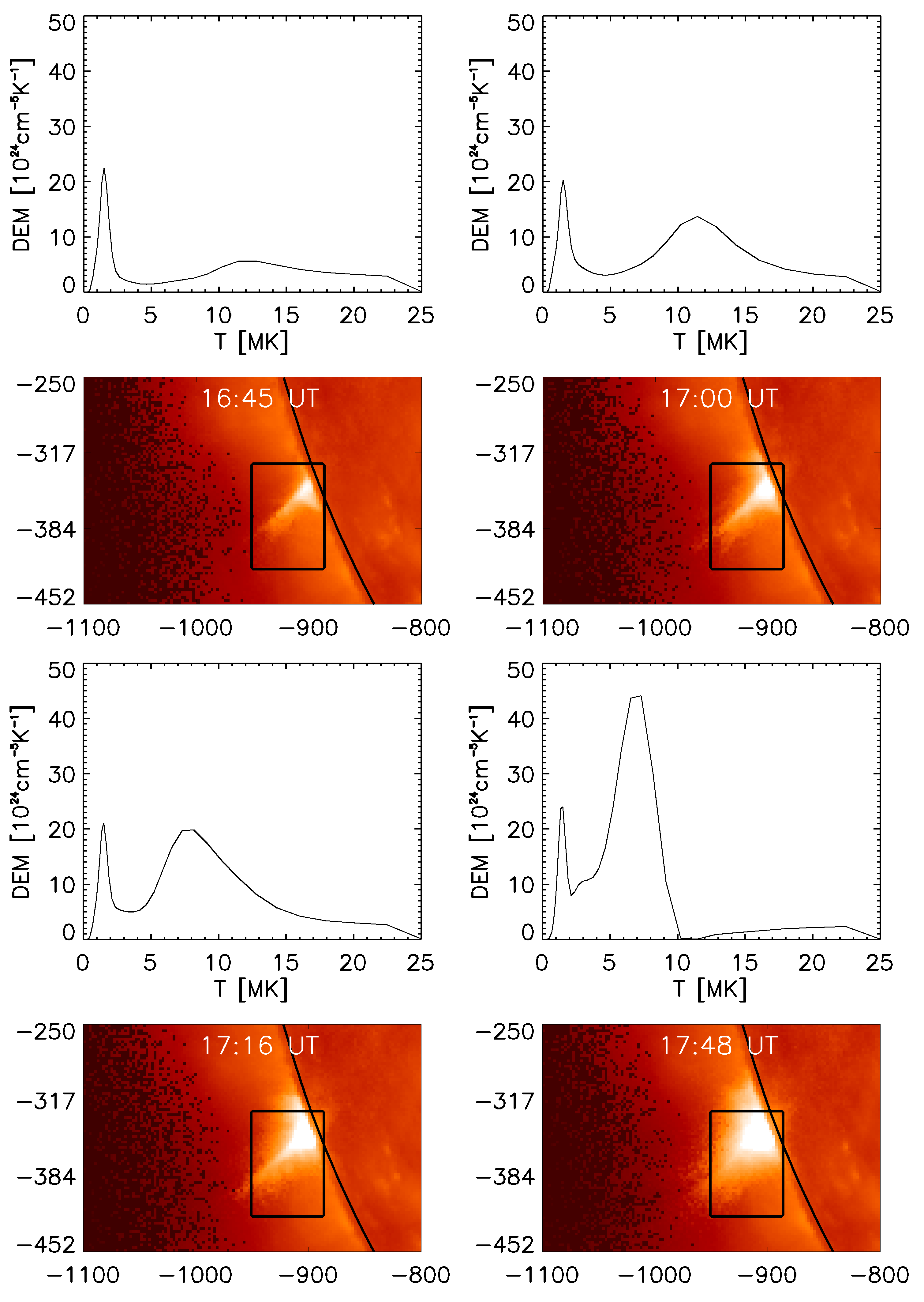}
\caption{DEM analysis of the off--limb flaring plasma visible from earth. DEM distributions summed over the rectangular area indicated in the total EM maps are shown for the end of the main flare phase (until 17:05\,UT) and the first 40\,minutes of the decay phase. An animation is included in the online supplementary material.}
\label{f-ArcadeEM}
\end{figure}
\par
After the energy release and particle acceleration, as evidenced by the STIX HXR light curve decreases, the plasma in the flaring region cooled down during the flare decay phase. This is illustrated in Fig. \ref{f-ArcadeEM}. The top row shows the DEM distribution summed over the regions indicated in maps of the total EM shown in the bottom row. Since most of the flaring plasma was occulted when viewed from Earth, the DEM increased significantly after 16:45\,UT as higher loops filled with plasma heated by the flare process appeared above the limb. The peak of the DEM distribution shifted from around 13\,MK at 17:00\,UT to 9\,MK at 17:16\,UT and further to 7\,MK at 17:48\,UT, as the plasma in these high flare loops cooled down after the HXR emission had ceased at 17:05\,UT.

\subsection{Spectral analysis and thermal diagnostics}
\begin{figure}
\centering
\includegraphics[width=0.43\textwidth]{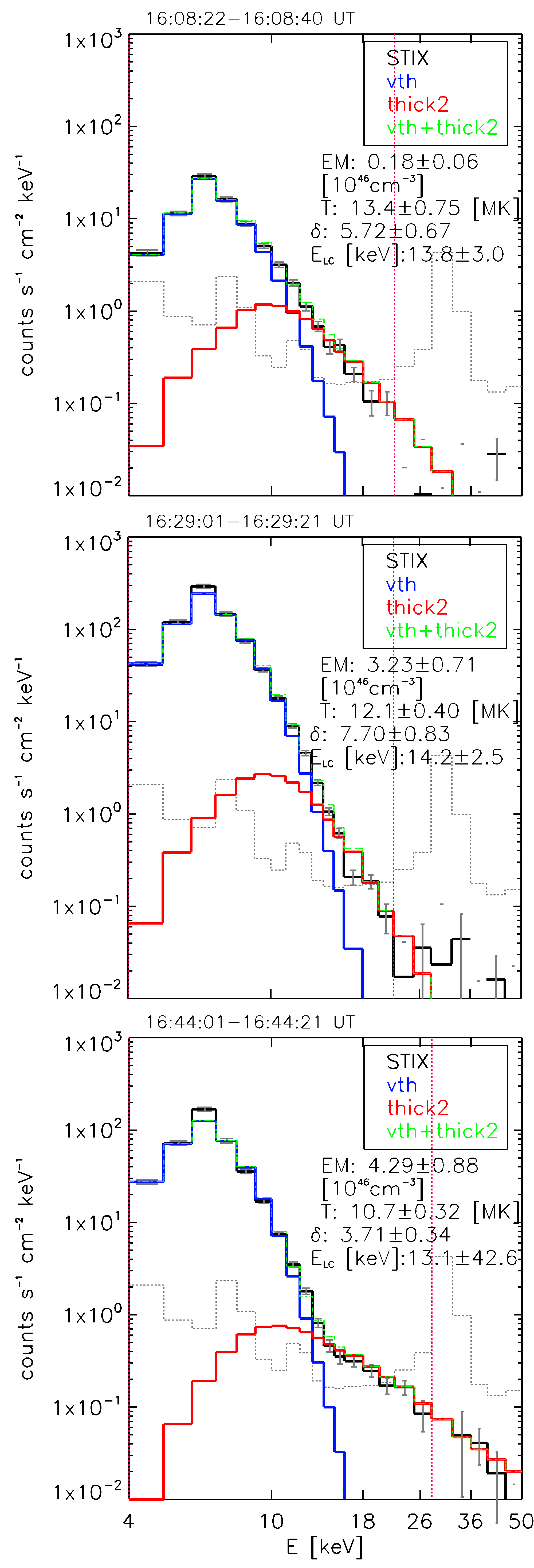}
\caption{STIX count spectra at three time intervals together with the functional fits. Black: Observed STIX count rate. Blue: Isothermal fit. Red: Thick target fit. Green: Sum of isothermal and thick target emission. Grey: Preflare background. Red vertical lines indicate the energy range considered for fitting.}
\label{f-spectra1}
\end{figure}
\begin{figure}
\centering
\includegraphics[width=0.48\textwidth]{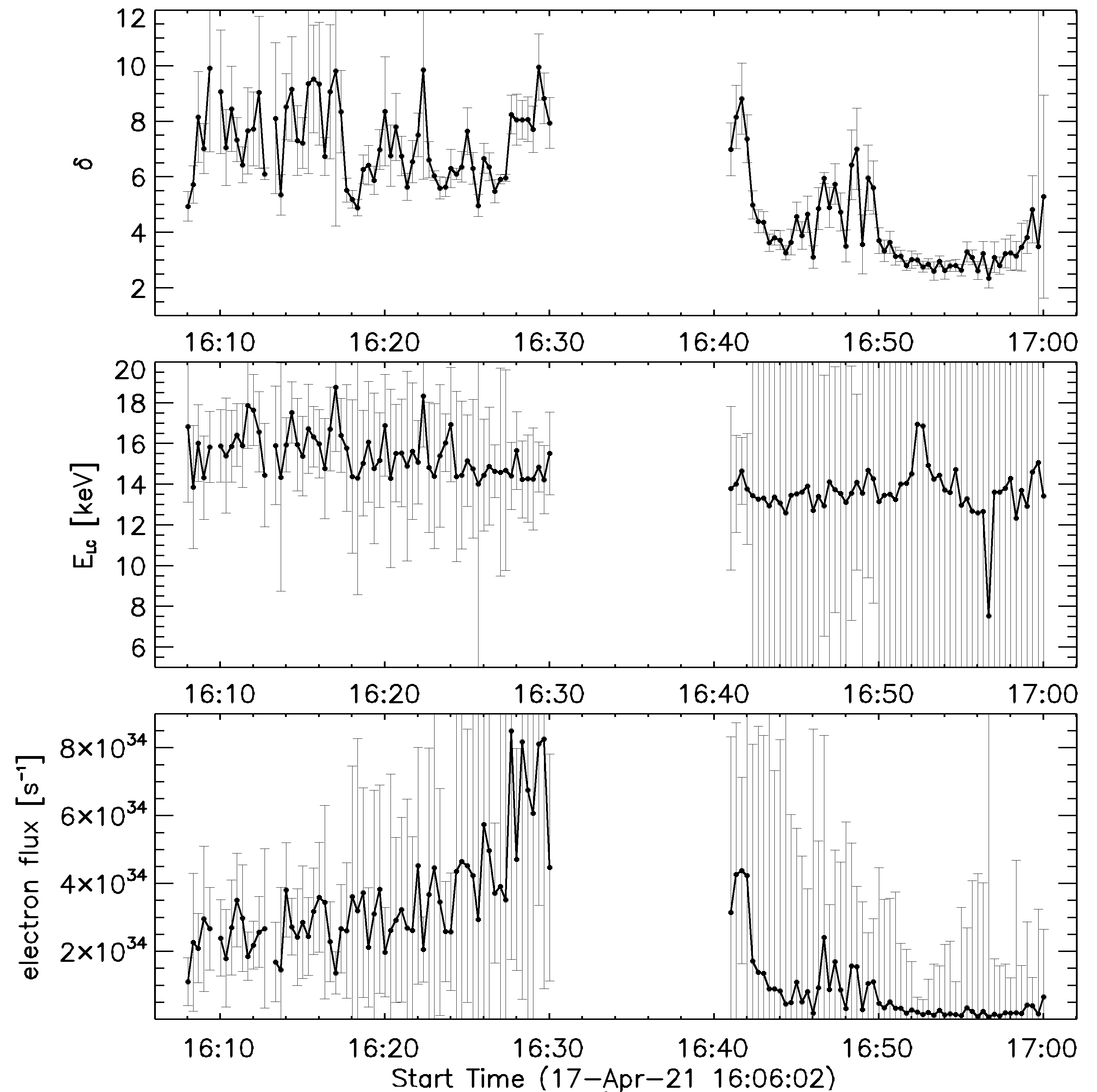}
\caption{Evolution of the thick target fit parameters to the STIX spectra. Top panel: Electron spectral index $\mathrm{\delta}$. Middle panel: Low energy cut off $\mathrm{E_{LC}}$. Bottom panel: Total integrated electron flux.}
\label{f-thickTargetEvolution}
\end{figure}
Figure \ref{f-spectra1} shows STIX spectra integrated over 20\,seconds as well as the functional isothermal and thick target fits. Four\,minutes after the beginning of the impulsive flare phase at 16:08\,UT, strong nonthermal emission was present in the form of multiple HXR bursts while the thermal component was relatively small, with a mean temperature of $T = 13.4$\,MK (top panel). At 16:29\,UT, the thermal emission measure had increased by an order of magnitude from $0.18$ to $3.23 \times \mathrm{10^{46} cm^{-3}}$.
The  bottom panel at 16:44\,UT shows the subsequent cooling of the thermal component and the tendency toward harder accelerated electron spectra during the later phase of the flare. Compared to the top panel at 16:08\,UT, the thermal component cools from 13.4 to 10.7\,MK, while the electron power law index of the thick target emission hardens from $\delta=5.7$ to $\delta=3.7$. Figure\ref{f-thickTargetEvolution} illustrates the evolution of the parameters from the thick target model fitted to the X-ray spectra over the course of the flare. The power law index $\delta$ shown in the top panel alternates between soft and harder nonthermal emission over the duration of the event. This evolution of the spectral index is known as soft-hard-soft in the literature (e.g., \citealt{parks1969,holman2011}). During the later phase of nonthermal emission after around 16:40\,UT, the total electron flux (bottom panel) became significantly reduced compared to the first phase of HXR emission, but the spectra were much harder (top panel). 
\par
\begin{figure}
\centering
\includegraphics[width=0.48\textwidth]{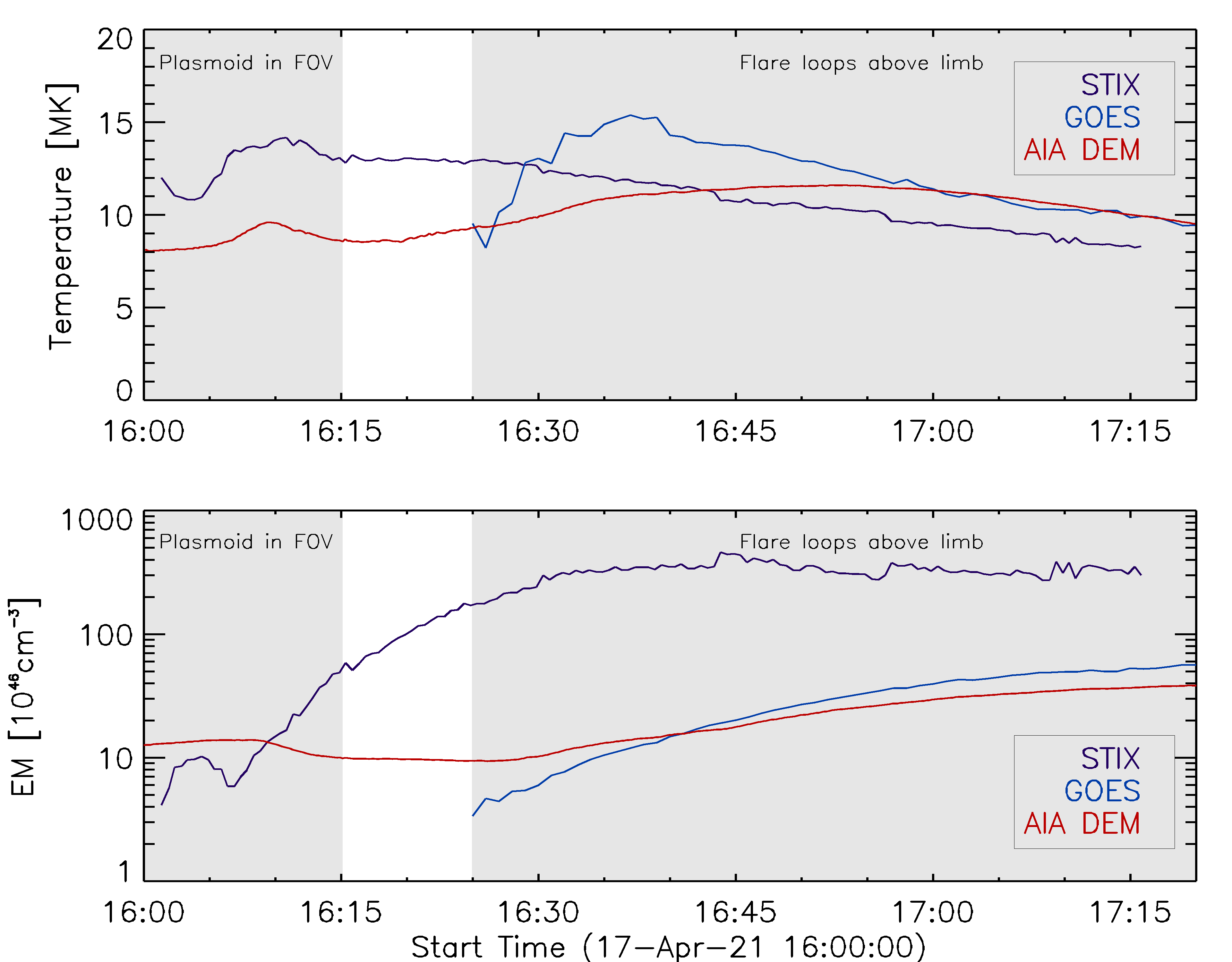}
\caption{Evolution of thermal plasma parameters. Top panel: Plasma temperatures over the entire flare region derived from different instruments. Bottom panel: Total EM for the same region. The first shaded region indicates the time span when the erupting plasmoid was detected with AIA. After the plasmoid left the FOV, most of the GOES and AIA emission was occulted behind the east limb. The second shaded region indicates when higher flare loops appeared above the limb, as seen from Earth, and the GOES signal was again strong enough to derive reliable results.}
\label{f-GOESDEMSTIXTEM}
\end{figure}

Figure \ref{f-GOESDEMSTIXTEM} shows the thermal plasma parameters $T$ and EM derived from AIA, GOES, and STIX spectral fitting. Direct comparisons are difficult not only because of the different instrument response functions but, in the present case in particular, also due to the occultation of the lower parts of the flare when seen from Earth. However, regions behind the limb from Earth's viewpoint are fully within the STIX FOV. The top panel shows the isothermal temperatures over the entire flare region deduced from STIX spectra and GOES as well as the emission-weighted temperature derived from the AIA DEM (from the region indicated in Fig. \ref{f-ArcadeEM}). The peak around 16:08\,UT for STIX and the DEM derived from AIA data corresponds to the hot plasmoid rising above the limb. In the EM evolution shown in the bottom panel, the AIA DEM shows an initial rise when the plasmoid appears from behind the limb for the SDO view. Afterwards, it reduces again until the hot flare loops appear in the SDO FOV (see also Fig.\ref{f-ArcadeEM}). For STIX, the EM of the thermal plasma rises gradually over the course of the event as the entire flaring region lies within its FOV. The gap at the start of the curves deduced from GOES is due to the insufficient signal when most of the flare was occulted. Although the 15--25\,keV\,STIX light curve shows two peaks around 16:44 and 16:52\,UT (Fig. \ref{f-STIXGOESLQ}), there is no corresponding increase in the EM or temperature from the STIX spectral fits (Fig. \ref{f-GOESDEMSTIXTEM}), but the EM stays at a high level during this later phase.
\par
\begin{figure}
\centering
\includegraphics[width=0.49\textwidth]{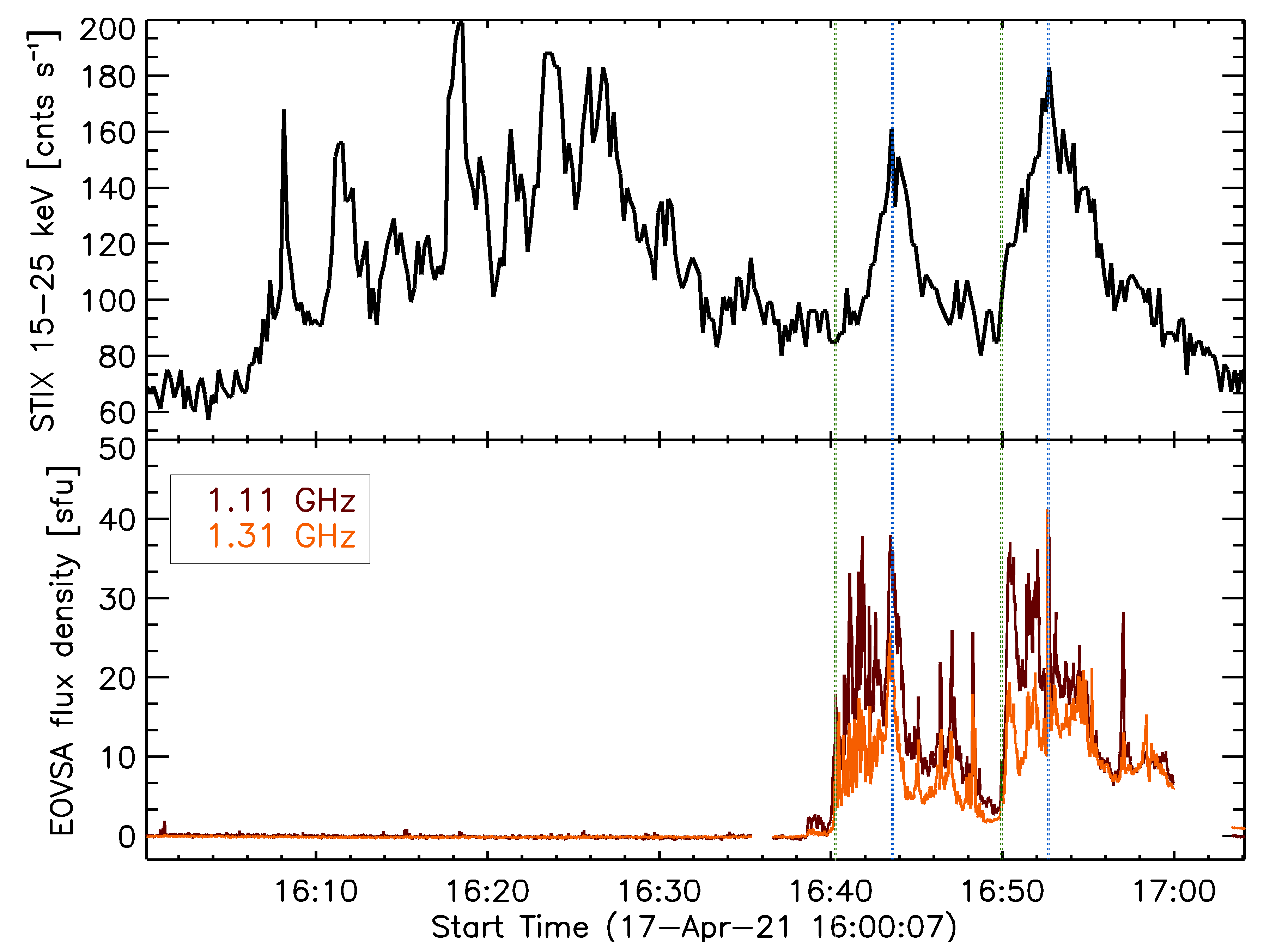}
\caption{EOVSA radio and STIX HXR light curves. Bottom panel: EOVSA flux curves at 1.11 and 1.31\,GHz. Top panel: STIX 15--25\,keV HXR light curve. The green dashed lines show the beginning of the radio emission for both peaks, and the blue lines indicate the STIX HXR peak times.}
\label{f-EOVSA_STIX}
\end{figure}

Figure \ref{f-EOVSA_STIX} shows EOVSA radio fluxes at 1.11 and 1.31 GHz along with the STIX 15--25\,keV light curve. There was no significant enhancement in the microwaves during the first phase of HXR emission until 16:40, although there was already enhanced GOES SXR emission starting at 16:22\,UT from plasma above the limb as seen from Earth (see Fig. \ref{f-STIXGOESLQ}). The radio emission must therefore be located at lower heights than the source of the GOES SXR flux and become visible from Earth only during the later phase. During the second HXR phase, the low frequency EOVSA channels showed two distinct episodes of emission starting at around 16:40 and 16:50\,UT that coincide with the increases in the STIX 15--25\,keV light curve related to the two late-phase HXR peaks (green lines in Fig. \ref{f-EOVSA_STIX}). The maxima of the microwave emission during the two episodes match the HXR peaks at 16:44 and 16:52\,UT.
\section{Discussion and conclusions}
Owing to the favorable spacecraft locations and the position of the active region under study, we were able to investigate the details of the eruption associated with a long duration GOES C7 class flare on 17 April 2021 from multiple viewpoints. The flaring region was located at $\mathrm{\phi}=-110^\circ$ and $\mathrm{\theta}=-18^\circ$ in Stonyhurst heliographic coordinates. The data utilized in this study are from spacecraft positioned such that the flaring region was occulted by 20$^\circ$ for SDO, observed on-disk at $\approx-57^\circ$long, $-11^\circ$lat by STEREO-A, and observed at $\approx-12^\circ$long, $-18^\circ$lat by the Solar Orbiter. 
\par
The flare showed several bursts of HXR emission in two separate phases for a little more than an hour, from 16:04\,UT to 17:05\,UT. The first phase is associated with a strong increase in emission from hot plasma and the acceleration peak of the associated CME, while the enhanced HXR emission during the second phase (after 16:30\,UT) is related to only small changes of the CME velocity and no significant further increase in the temperature or emission measure of the thermal flare plasma.
\par
During the first phase, the associated CME was found to be initiated by a rising plasmoid, which was observed by AIA and showed a distinct two-phase evolution, with a slow rise and impulsive acceleration. After it appeared above the limb, as seen from Earth, it gradually rose to $\mathrm{0.073 \pm 0.004}$\,$\mathrm{R_{\odot}}$\,($\mathrm{50.8 \pm 2.8}$\,$\mathrm{Mm}$) over the solar limb in the time from 15:59 to 16:08\,UT. The DEM analysis showed that the rising plasmoid was significantly hotter than the quiet sun plasma, with the main contribution centered around 14\,MK as soon as it was detected above the limb by AIA.
\par
Cotemporal STIX X-ray observations taken from a vantage point where the event was observed against the disk show the first HXR emission starting at around 16:04\,UT. When the plasmoid started accelerating at 16:08\,UT, the HXR emission increased significantly by an order of magnitude. Flare kernels appeared on-disk in the STEREO-A EUV filters and were followed by a cusp above the limb, as seen by AIA 131\,\AA\,, indicating magnetic reconnection taking place in a large-scale current sheet beneath the erupting structure. 
\par
Following the onset of the eruption, we observed further signatures of magnetic reconnection behind the CME, such as additional smaller-scale outflows, the appearance of another cusp structure, and supra-arcade downflows in the high-temperature AIA 131\,\AA\,channel as well as distinct nonthermal HXR bursts in the STIX light curves.
The CME acceleration peaked at 16:21\,UT, when it reached $\mathrm{395 \pm 42}$\,$\mathrm{m/s^{2}}$ at a distance of 0.51\,$\mathrm{R_{\odot}}$ from the solar limb. The disappearance of the reconnection signatures coincided with the end of the CME acceleration around 16:50\,UT, when the peak CME velocity of $\mathrm{802 \pm 44}$\,$\mathrm{km/s}$ was reached 2.4\,$\mathrm{R_{\odot}}$ from the limb. After the acceleration had ceased, the CME gradually decelerated to $\mathrm{634 \pm 82}$\,$\mathrm{km/s}$ until 18:23\,UT, when it could not be reliably tracked in SECCHI/COR2 images anymore. This three-part evolution of the CME kinematics and the correlation of the CME acceleration profile with the flare energy release and particle acceleration, as diagnosed by the HXR emission, is in line with previous studies (see e.g., \citealt{zhang2001,temmer2008,temmer2010,b-s2012,veronig2018}). The clear indications of reconnection behind the CME during the acceleration phase support the standard eruptive flare scenario where reconnection is expected to occur in the large-scale current sheet formed behind the rising core field (see e.g., review by \citealt{green2018}). Our observation of a second HXR energy release phase, which is not related to further CME acceleration, contrasts with the study of a fast CME by \cite{gou2020}, where two peaks in the CME acceleration profile that were associated with two phases in the flare HXR emission could be reconstructed.
\par
Due to occultation, EOVSA shows decimeter radio emission only during the second phase of HXR energy release. The episodes of microwave emission correspond to the two late phase HXR peaks in their onset and peak times. We interpret the detected radio signal as coherent plasma emission due to its high spectral and temporal variability. This corroborates the detection of nonthermal electrons.
\par
After the main CME acceleration and the associated HXR peaks subsided at 16:30\,UT, the STIX X-ray spectra show a hardening, with the electron spectral index $\delta$ changing from approximately seven   to approximately four. The nonthermal emission during this later phase of HXR emission did not lead to a significant increase in thermal plasma (see isothermal spectral fits in Fig. \ref{f-GOESDEMSTIXTEM}) because the total flux of nonthermal electrons was reduced compared to earlier HXR peaks and the flatter electron spectrum (Fig. \ref{f-thickTargetEvolution}) contains fewer low-energy nonthermal electrons, which are most efficient at producing chromospheric evaporation \citep{reep2015}.
\par
While the observation of flare kernels, the appearance of bright EUV loops, and the increase in EM during the first phase of HXR emission supports the standard chromospheric evaporation scenario, the EM during the second HXR phase remains at a high level instead of increasing further, which would be expected based on the Neupert effect. We conclude that the energy deposited into the flare loops by chromospheric evaporation during the second HXR phase is smaller than the instantaneous energy losses by thermal conduction and radiation \citep{veronig2005}, which is supported by the spectral fitting results showing a continuous decrease of the plasma temperature and a constant EM, even during phases of new HXR emission (Fig. \ref{f-GOESDEMSTIXTEM}). A lack of chromospheric evaporation during late HXR peaks in a solar flare has been previously reported for an X-class event by \citealt{warmuth2009}, who proposed a different acceleration process taking over or a change in the characteristic parameters of the initial accelerator as possible mechanisms. For the event under study, we can neither confirm nor rule out such a scenario due to the lack of magnetic field observations of the active region.

 \begin{acknowledgements}
JS and AMV acknowledge the Austrian Science Fund (FWF): I4555-N. AFB is supported by the Swiss National Science Foundation Grant 200021L\_189180 for STIX. Solar Orbiter is a space mission of international collaboration between ESA and NASA, operated by ESA. The STIX instrument is an international collaboration between Switzerland, Poland, France, Czech Republic, Germany, Austria, Ireland, and Italy.
\end{acknowledgements}

%
%

\bibliographystyle{aa} 
\bibliography{biblio.bib} 

\end{document}